%% file: main.tex
\documentclass[conference]{IEEEtran}
\IEEEoverridecommandlockouts
\usepackage{cite}
\usepackage{amsmath,amssymb,amsfonts}
\usepackage{algorithmic}
\usepackage{graphicx}
\usepackage{textcomp}
\usepackage{xcolor}
\def\BibTeX{{\rm B\kern-.05em{\sc i\kern-.025em b}\kern-.08em
    T\kern-.1667em\lower.7ex\hbox{E}\kern-.125emX}}
    
\usepackage{cite}
\usepackage{amsfonts}
\usepackage{amsmath}
\usepackage{amssymb}
\usepackage{algorithm}
\usepackage{algorithmic}
\usepackage{graphicx}
\usepackage{textcomp}
\usepackage{xcolor}
\usepackage{hyperref}
\hypersetup{
 colorlinks=true,
 citecolor=black,
 linkcolor=black,
 urlcolor=blue}
 
\usepackage{listings}
\usepackage{float}
\usepackage[mode=buildnew]{standalone}
\usepackage{tikz,pgfplots}
\usetikzlibrary{automata,arrows}
\usetikzlibrary{calc,trees,positioning,arrows,chains,shapes}
\usetikzlibrary{shapes.geometric}
\usetikzlibrary{shapes.multipart}
\definecolor{darkgreen}{rgb}{0,0.6,0}\usepackage{tikz,pgfplots}
\usetikzlibrary{automata,arrows}
\usetikzlibrary{calc,trees,positioning,arrows,chains,shapes}
\usetikzlibrary{shapes.geometric}
\usetikzlibrary{shapes.multipart}
\definecolor{darkgreen}{rgb}{0,0.6,0}
\usepackage{enumitem}
\setlist[description]{leftmargin=0em,labelindent=\parindent,font=\normalfont\itshape}

\newfloat{lstfloat}{htbp}{lop}
\floatname{lstfloat}{Listing}

\lstdefinestyle{Python}{
    language        = Python,
    basicstyle      = \scriptsize\ttfamily,
    keywordstyle    = \color{blue},
    keywordstyle    = [2] \color{teal}, 
    stringstyle     = \textcolor{green},
    commentstyle    = \color{red}\ttfamily,
    emphstyle       = \color{blue},
    morekeywords    ={sin,cos}
}

\DeclareMathOperator{\Tr}{Tr}

\def\BibTeX{{\rm B\kern-.05em{\sc i\kern-.025em b}\kern-.08em
    T\kern-.1667em\lower.7ex\hbox{E}\kern-.125emX}}
\pgfplotsset{compat=1.17}

\begin{document}

\title{Memory-Efficient Differentiable Programming for Quantum Optimal Control of Discrete Lattices
\thanks{ This material is based upon work supported by the U.S. Department of Energy, Office of Science, Office of Advanced Scientific Computing Research, under the Accelerated Research in Quantum Computing and Applied Mathematics programs, under contract DE-AC02-06CH11357, and by the National Science Foundation Mathematical Sciences Graduate Internship. We gratefully acknowledge the computing resources provided on Bebop and Swing, a high-performance computing cluster operated by the Laboratory Computing Resource Center at Argonne National Laboratory.}
}

\makeatletter 
\newcommand{\linebreakand}{%
  \end{@IEEEauthorhalign}
  \hfill\mbox{}\par
  \mbox{}\hfill\begin{@IEEEauthorhalign}
}
\makeatother 

\author{\IEEEauthorblockN{Xian Wang}
\IEEEauthorblockA{\textit{
University of California, Riverside} \\
xwang056@ucr.edu}
\and
\IEEEauthorblockN{Paul Kairys}
\IEEEauthorblockA{\textit{Argonne National Laboratory} \\
pkairys@anl.gov}
\and
\IEEEauthorblockN{Sri Hari Krishna Narayanan}
\IEEEauthorblockA{\textit{Argonne National Laboratory} \\
snarayan@anl.gov}
\and \linebreakand
\IEEEauthorblockN{Jan H\"uckelheim}
\IEEEauthorblockA{\textit{Argonne National Laboratory} \\
jhueckelheim@anl.gov}
\and 
\IEEEauthorblockN{Paul Hovland}
\IEEEauthorblockA{\textit{Argonne National Laboratory} \\
hovland@mcs.anl.gov}
}

\maketitle

\begin{abstract}
Quantum optimal control problems are typically solved by gradient-based algorithms such as GRAPE, which suffer from exponential growth in storage with increasing number of qubits and linear growth in memory requirements with increasing number of time steps. Employing QOC for discrete lattices reveals that these memory requirements are a barrier for simulating large models or long time spans. We employ a nonstandard differentiable programming approach that significantly reduces the memory requirements at the cost of a reasonable amount of recomputation. The approach exploits invertibility properties of the unitary matrices to reverse the computation during back-propagation. We utilize QOC software written in the differentiable programming framework JAX that implements this approach, and demonstrate its effectiveness for lattice gauge theory.
\end{abstract}


\section{Introduction}

Quantum control allows systems that obey the laws of quantum mechanics must be manipulated to create desired behaviors. The application of external electromagnetic fields or force affects dynamical processes at the atomic or molecular scale~\cite{Werschnik_2007}.
Quantum optimal control (QOC) approaches determine the external fields and forces to achieve a task in a quantum system in the best way possible\cite{kairys_parametrized,holland2020,lysne_small}. In particular, QOC can be used to achieve state preparation and gate synthesis.


One of the computational advantages of quantum information processing is realized through the efficient simulation of quantum mechanical effects \cite{georgescu2014quantum, lloyd1996universal}. This potential impact is considerable within the fields of condensed matter and particle physics where the simulation of large quantum systems is critical for scientific discovery. In particular, the study of lattice gauge theories (LGT) provides significant insight into fundamental and emergent physics and is a critical application for quantum simulation \cite{snowmass,MARCOS2014634,martinez2016real,banuls2020simulating}.
\par
One potential route to achieving high-fidelity quantum simulation is through the use of QOC. In this application, QOC provides a compilation of the desired unitary process $U_{target}$ onto a set of analog device controls $\vec{\alpha}$. Using optimal control to implement the simulation is advantageous for two reasons. First, by reducing the device time needed to implement a specific unitary process one achieves higher fidelity due to reduced decoherence. Second, decomposing the desired unitary process into a locally-optimal quantum gate set accrues an approximation error and an optimal control route avoids this by compiling the desired unitary directly.
\par
One of the major downsides of QOC for quantum simulation is due to the need to accurately model the parameterized device evolution $U_{device}(\vec{\alpha})$. While in principle this optimization can be accomplished without additional information, accessing the derivative information, i.e. $\partial_{\alpha_i} U_{device}(\vec{\alpha})$ can dramatically accelerate the optimization protocol but comes with additional computational overhead. Our work assesses how this burden can be lifted by using memory-efficient differentiable programming strategies and applies these strategies to simulations of LGTs on superconducting quantum computers.
\par

We follow the QOC model of ~\cite{PhysRevA.95.042318}. Given a Hamiltonian $H_0$, an initial state $|\psi_0\rangle$, and a set of control operators
$H_1, H_2, \ldots H_m$, one seeks to determine, for a sequence of time steps
$t_0, t_1, \ldots, t_N$, a set of control fields $g_i(t)$ such that

\begin{eqnarray}
\mathbb{H}_t & = & H_0 + \sum_{i=1}^{m}g_i(t)H_i \label{evovlveschrodingersicrete1}\\
U_t & = & e^{-i\mathbb{H}_t\Delta t}\label{evovlveschrodingersicrete2}\\
K_t & = & U_{t}U_{t-1}U_{t-2} \ldots U_{1}U_{0}\\
|\psi_t\rangle & = & K_t|\psi_0\rangle. \label{evovlveschrodingersicrete4}
\end{eqnarray}

One possible objective is to minimize the trace distance between $K_N$ and a target quantum gate $K_T$:
\begin{eqnarray}
F_0 = 1 - |\Tr(K^{\dagger}_TK_N)/D|^2,  
\end{eqnarray}
where $D$ is the Hilbert space dimension.
\par
In this work we approach QOC using the gradient ascent pulse engineering (GRAPE) algorithm~\cite{PhysRevA.63.032308}, as shown in Algorithm~\ref{algo:grape}. The algorithm requires derivative terms $\frac{ \partial \rho_t \lambda_t }{\partial g_i(t)}$ that can be accurately calculated using automatic differentiation (AD or autodiff)~\cite{PhysRevA.95.042318}, a well-known technique for obtaining derivatives and gradients of numerical functions~\cite{Griewank2008EDP,Naumann_book,baydin2015automatic}.

\begin{algorithm}
	\caption{Pseudocode for the GRAPE algorithm.}
	\label{algo:grape}
	\begin{algorithmic}
		\STATE Guess initial controls $g_i(t)$.
		\REPEAT
		    \STATE Starting from $H_0$, calculate \\ { \quad\quad\quad $\rho_t=U_tU_{t-1}\ldots U_1 H_0 U_1^\dagger \ldots U_{t-1}^\dagger U_{t}^\dagger$}.
		    \STATE Starting from $\lambda_N =K_T$, calculate \\ { \quad\quad\quad $\lambda_t=U_{t+1}^\dagger \ldots U_N^\dagger K_T U_N \ldots U_{t}$}.
		    \STATE Evaluate $\frac{ \partial \rho_t \lambda_t }{\partial g_i(t)}$ 
	            \STATE Update the $m \times N$ control amplitudes: \\ \quad\quad\quad $g_i(t) \rightarrow g_i(t)+\epsilon \frac{ \partial \rho_t \lambda_t }{\partial g_i(t)}$
		\UNTIL{ $\Tr{(K_T^{\dagger}K_N)} <$ threshold}
		\STATE \textbf{return} $g_i(t)$
	\end{algorithmic}
\end{algorithm}

For computations with many input parameters, it is often most efficient to use the so-called \emph{reverse mode} of AD, which has been popularized as \emph{back-propagation} in machine learning. Reverse mode AD computes the derivatives of a function's output with respect to its inputs by tracing sensitivities backwards through the computational graph after the original computation is completed. Since QOC has a large number of input parameters and few outputs (only the cost function(s)), reverse mode AD is a promising approach.

However, reverse mode AD requires that certain intermediate states of the original computation are available during the derivative computation. In the case of QOC, storing such values in order to re-use them during the derivative computation results in additional memory usage that is exponentially 
proportional to the number of qubits as well as proportional to the number of time steps, severely limiting the system size and duration that can be simulated on classical computers. Our previous work~\cite{10.1007/978-3-031-08760-8_11} introduced non-standard approaches for reducing the memory requirements of QOC through recomputation or by exploiting reversibility, and we apply these approaches to lattice gauge theory in this work.

There exist several implementations of quantum control. {\tt QuantumControl.jl} and its subpackages {\tt GRAPE.jl} and 
{\tt Krotov.jl} provide a Julia framework for quantum optimal 
control. {\tt GRAPE.jl} is an implementation of (second-order) 
GRAPE extended with automatic differentiation. {\tt GRAPE.jl} optimizes its memory utilization and achieves low runtime using a technique that combines analytical derivatives with naive automatic differentiation. Their approach is suitable for both open as well as closed quantum systems. 

{\tt QuTiP} is open-source software for simulating the dynamics of open quantum systems in Python and utilizes the Numpy, Scipy, and Cython numerical packages~\cite{JOHANSSON20131234}. For the derivative-based optimal control it uses the GRAPE algorithm, where control pulses are piece-wise constant functions~\cite{Li2022PulselevelNQ}. {\tt QuTiP} also provides the derivative-free CRAB algorithm. {\tt Krotov} is a Python library that supports optimal control in closed and open systems~\cite{10.21468/SciPostPhys.7.6.080}.

Classical differentiable programming frameworks like JAX provide autodiff capabilities.  One approach to differentiable programming for quantum control uses reinforcement learning. Here, a control agent is represented as a neural network that maps the state of the system at a given time to a control pulse. The parameters of this agent are optimized via gradient information obtained by direct differentiation through both the neural network and the differential equations of the system~\cite{Sch_fer_2020,murphy2019}.

The rest of the paper is organized as follows.
Section~\ref{sec:approach} presents our differentiable programming approach for reducing the memory requirements of QOC. Section~\ref{sec:lgt} discusses LGT. 
Simulation results are presented in Section~\ref{sec:results}. Section~\ref{sec:conclusion} concludes the paper.

\section{Approach}
\label{sec:approach}

We apply the three ``advanced'' automatic differentiation approaches presented in~\cite{10.1007/978-3-031-08760-8_11}, which we summarize in this section, as well as a simpler ``naive'' approach. All four approaches are used to restore intermediate values of the computation when they are needed during the subsequent derivative computation.

\begin{description}
\item[Naive Approach (Store-All)]
retains in memory all intermediate values that will be needed for the derivative computation, and is the default in JAX and many other frameworks and AD tools such as PyTorch, Tapenade, etc.
\item [Periodic Checkpointing]
is an AD technique that stores selected intermediate values in memory so that they can later be loaded during the subsequent derivative computation. Values that have not been stored will instead be recomputed, by restarting parts of the computation from the nearest available earlier  state. Periodic checkpointing is a sub-optimal approach but is straightforward to implement. To compute the derivative of an interval, the intermediate states are recomputed and kept in memory for the duration of the derivative computation of that interval. The checkpointing approach reduces the overall memory consumption compared to a store-all approach, at the cost of some recomputation.
\item [Reversibility]
exploits the fact that the inverse of unitary matrices is their conjugate transpose, which can be computed cheaply and accurately. The use of the inverse allows computing $K_{t-1}$ from $K_{t}$ and $\psi_{t-1}$ from $\psi_{t}$.
\begin{eqnarray}
K_t & = & U_{t}U_{t-1}U_{t-2} \ldots U_{1}U_{0}\\
\label{eq:useinverse}
K_{t-1}& = & U_{t}^\dagger K_t\\
\label{eq:useinversestate}
|\psi_{t-1}\rangle& = & U_{t}^\dagger|\psi_{t}\rangle
\end{eqnarray}
Thus, one does not have to store any of the $K_t$ matrices required to compute the adjoint of
a time step. Additionally, reversibility allows a further memory reduction by avoiding the storage of $U_t$, and recomputing it from the $g_i(t)$ control values instead. While this drastically reduces memory consumption and recomputation cost compared to checkpointing approaches, it potentially incurs roundoff errors during the inversion of long time-step sequences.
\item[Checkpointing with Reversibility]
is the third advanced approach, which combines checkpointing and reversibility to combine the accuracy of checkpointing with the efficiency of reversibility approaches. Checkpoints are stored at regular intervals as in the first approach, but the intermediate states within each interval are obtained by reversing the trajectory backwards from the final state of the interval.
\end{description}

The approaches are implemented in JAX, a differentiable programming framework that
can automatically differentiate native Python and NumPy functions~\cite{jax2018github}. It can differentiate through loops, branches, recursion, and closures, and it can take derivatives of derivatives of derivatives. It supports reverse-mode differentiation (a.k.a. backpropagation) via {\tt grad} as well as forward-mode differentiation, and the two can be composed arbitrarily to any order.

We have used JAX's {\tt jax.custom\_vjp} feature to implement the three advanced approaches. Using the feature, one can  provide derivatives for a portion of the computation instead of relying on JAX's standard approach. Listing~\ref{lst:custom_derivatives} shows how the derivative for {\tt f(x,y)} can be computed analytically and used in the overall derivative computation.

\begin{lstfloat}
\begin{lstlisting}[style=Python]
@jax.custom_vjp
def f(x, y):
  return jnp.sin(x) * y

def f_fwd(x, y):
  return f(x, y), (jnp.cos(x), jnp.sin(x), y)

def f_bwd(res, g):
  cos_x, sin_x, y = res
  return (cos_x * g * y, sin_x * g)

f.defvjp(f_fwd, f_bwd)
\end{lstlisting}
\caption{Custom derivatives in JAX.}
\label{lst:custom_derivatives}
\end{lstfloat}

\section{Benchmark Application}
\label{sec:lgt}
We will restrict our discussion to the simulation of qubit systems but wish to emphasize that our analysis and methods are also applicable to arbitrary quantum systems. To assess the reduced memory footprint that the combination of checkpointing and reversibility provides, we have explored the task of quantum simulation lattice gauge theories using optimal control. 
\par
In this application context one specifies a model Hamiltonian $H_\text{model}$ and device Hamiltonian $H_\text{device}(\vec{\alpha},t)$ and uses optimization to determine a set of controls $\vec{\alpha}\*$ that yields a device evolution $U_\text{device}$ close to the desired model evolution $U_\text{model}$ \cite{kairys_parametrized, lysne_small}. This is often difficult because the simulation of $n$ qubits requires computing and storing operators defined on a Hilbert space with dimension $2^n$, growing exponentially large with increasing system size. To alleviate this, one typically decomposes the model evolution from a single global unitary defined on $n$ qubits to a product of unitary evolutions with support on only $m$ qubits through the Lie-Trotter decomposition \cite{lloyd1996universal,childs2021theory, kairys_parametrized}. \par
Commonly referred to as Trotterization, applying the Lie-Trotter decomposition only approximates the global unitary to some error. Furthermore, when choosing $m$ to be small (which reduces classical computational overhead by limiting the classical simulation to $m$ qubits) this will tend to yield larger Trotter error and requires deeper quantum circuits to mitigate \cite{childs2021theory}. Thus increasing $m$ as much as possible will help to mitigate errors due to approximations and reduce circuit depth, limiting errors due to decoherence. Using methods of checkpointing and reversibility enable optimal control studies of larger quantum systems and therefore could enable more accurate and efficient quantum simulations.
\par
\begin{figure}
    \centering
    \includegraphics[scale=0.5]{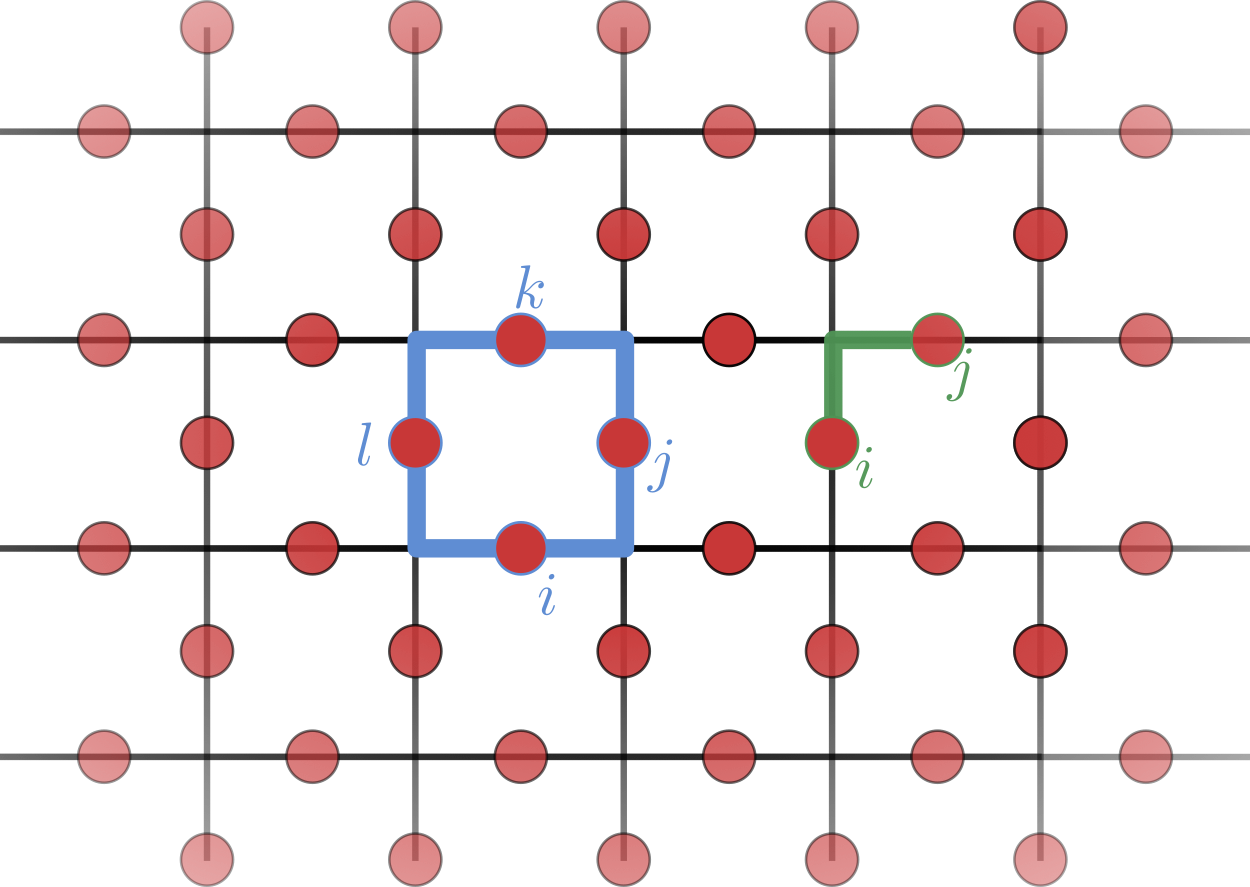}
    \caption{The definition of a $U(1)$ lattice gauge theory on a square lattice with spin-$1/2$ particles as defined in Ref.~\cite{MARCOS2014634}. The spin-$1/2$ particles (respectively, qubits) are denoted as circles positioned on the edges of the lattice. The global Hamiltonian of the lattice is defined as the sum of ``plaquette'' operators and ``corner'' operators. Square operators have non-identity support on four spins within a single square (highlighted in blue) and corner operators have non-identity support on two spins on each corner of the lattice (highlighted in green).}
    \label{fig:LGT}
\end{figure}
\par
\begin{figure*}
\begin{center}
    \includegraphics[width=0.8\textwidth]{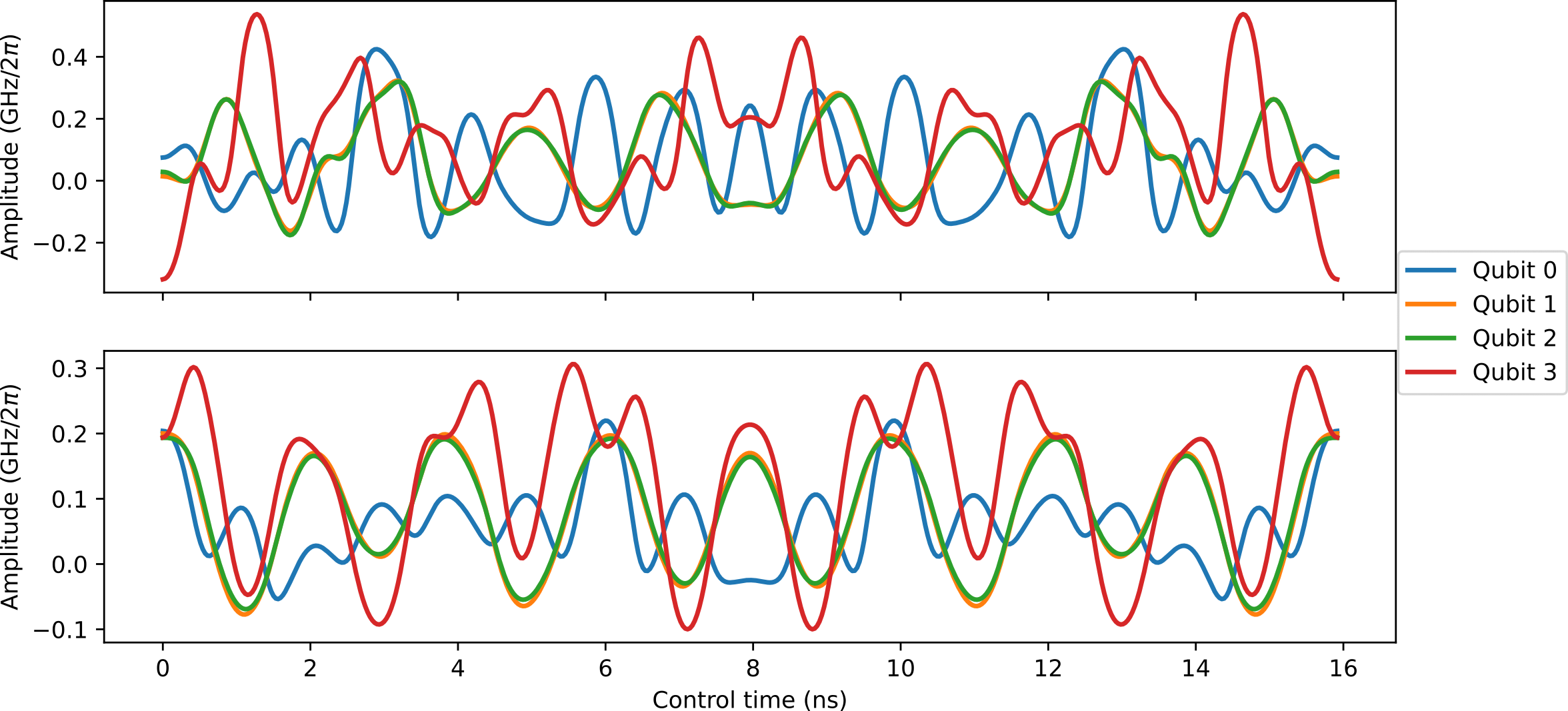}
    \end{center}
    \caption{Two sets of identified controls after 1000 iterations of optimization for a lattice with 4 qubits. The gradients during optimization were calculated via the Reversibility method. Both figures use the same device Hamiltonian and assume $t/q = 0.01$ ns. The top figure visualizes optimized controls that generate $U_P(t/q)$ for $J = 1$ with infidelity of $F \approx 8.3\times 10^{-6}$. The bottom figure visualizes optimized controls that generate $U_C(t/q)$ for $V = 1$ with infidelity of $F \approx 2.0\times 10^{-5}$.}
    \label{fig:optimal_pulses}
\end{figure*}
\par
One application instance in which $m$ is large is the simulation of quantum systems exhibiting non-local interactions. A family of systems which exhibit these non-local interactions are found within the class of LGTs, which describe both fundamental and emergent physics and are a prime application for quantum simulation \cite{banuls2020simulating}. 
\par
We choose to focus on a 2-dimensional $U(1)$ LGT model that has been explored in the context of analog simulation with superconducting circuits in Ref.~\cite{MARCOS2014634}. In that work the LGT is described by the model Hamiltonian:

\begin{align}\label{eq:model_ham}
    H_\text{model} &= -J H_P + V H_C
\end{align}

where spins (respectively, qubits) are defined on the edges of a square lattice shown in Figure~\ref{fig:LGT} and $H_P$ denotes a Hamiltonian of ``plaquette" terms involving 4-local operators defined on each square of the lattice (highlighted in blue in the Figure~\ref{fig:LGT})

\begin{equation}
    H_P = \sum_{\square} ( S_+^{(i)} S_-^{(j)} S_+^{(k)} S_-^{(l)} + h.c.)
\end{equation}

and $H_C$ denotes a Hamiltonian of ``corner" terms involving 2-local operators defined on each corner of the lattice (highlighted in green in Figure~\ref{fig:LGT}):

\begin{equation}
    H_C = \sum_{\ulcorner} S_z^{(i)} S_z^{(j)}.
\end{equation}

The operators $S_{\pm}^{(j)} = S_x^{(j)}\pm i S_y^{(j)}$ denotes the qubit $i$ raising and lowering operator, $S_x^{(j)},S_y^{(j)},S_z^{(j)}$ are the qubit spin matrices, and h.c. denotes Hermitian conjugate. The notation $\sum_\square$ represents the sum over all square plaquettes on the lattice, and $\sum_{\ulcorner}$ is the sum over pairs of qubits in each corner which share a vertex \cite{MARCOS2014634}.
\par
One can approximate the global time evolution operator $U_\text{model}(t) = \exp(-\frac{i t}{\hbar} H_\text{model})$ as a product of local operators via Trotterization:

\begin{align}
    U_\text{model}(t) = \lim_{q \rightarrow \infty} \bigg[ U_{P}\bigg(\frac{t}{q}\bigg) \cdot U_{C}\bigg(\frac{t}{q}\bigg)\bigg]^q
\end{align}

where $U_P = \exp[\frac{it}{q\hbar} J H_P]$ and $U_C = \exp[-\frac{it}{q\hbar} V H_C]$ are the time evolution operators under only the plaquette and corner Hamiltonian and equality holds in the limit of $q$. Typically, one truncates this limit at finite $q$ which yields a $q$th order approximation to the unitary dynamics $U^q_\text{model}(t)$ with error $\Delta U_\text{model}^q(t)= U_\text{model}(t) - U^q_\text{model}(t)$ scaling polynomially in $t/q$:

\begin{align}
    \Delta U_\text{model}^q(t) = \frac{t^2}{2q} \sum_{l > m =1}^M [H_l,H_m] + \mathcal{O}\bigg(\frac{t^3}{q^2}\bigg)
\end{align}

where the sum of commutators is over every term in the Hamiltonian of Eq.~\ref{eq:model_ham}. 
\par
The four-body interaction terms given by $S_+^{(i)} S_-^{(j)} S_+^{(k)} S_-^{(l)}$ are known as ``ring exchange'' interactions and represent a unique non-local operator \cite{MARCOS2014634}.
This model Hamiltonian exhibits a number of interesting properties such as emergent excitations in the ground state and a quantum phase transition in the ratio of $J/V$ \cite{MARCOS2014634}.
This model Hamiltonian was chosen as an application in which the memory advantages of checkpointing and reversibility could provide meaningful utility.
Specifying the model Hamiltonian is only part of the example application. We also need to specify a Hamiltonian that models the assumed quantum device on which the simulation will be implemented. In this work we choose a device Hamiltonian derived from a two-dimensional array of coupled superconducting transmons such as those used to demonstrate quantum supremacy in 2019 \cite{arute2019quantum}.
\par

In this work we approximate the system as a set of coupled qubits, neglecting higher energy levels \cite{krantz2019quantum}. When a set of coupled transmons are tuned into resonance with one another their effective Hamiltonian can be described as:

\begin{align}
    H(\vec{\alpha},t) &= \sum_{i} \gamma_{i}(\vec{\alpha},t) S_x^{(i)}+ \sum_{\langle i,j \rangle} g (S_x^{(i)}S_x^{(j)} + S_y^{(i)}S_y^{(j)})
\end{align}

where $\sum_{\langle i,j \rangle}$ is the sum over all neighboring qubits on a square lattice, $g = -20 \times 2\pi$~MHz is a typical coupling strength between transmons, and $\gamma_i(\vec{\alpha},t)$ are the time-dependent microwave control envelope functions modulated in resonance with the transmon frequencies \cite{krantz2019quantum,arute2019quantum}.
\par
Thus the optimal control task is to determine a set of controls $\vec{\alpha}$ that minimizes the infidelity as defined by

\begin{align}
    F(U_\text{model},U_\text{device}) = \frac{|\Tr(U_\text{model}^\dagger U_\text{device})|^2}{D^2}
    \label{eq:infidelity}
\end{align}

where $D$ is the dimension of the Hilbert space on which $U_\text{model}$ is defined \cite{d2021introduction}. 

\section{Experimental Results}
\label{sec:results}
We first provide a validation that the reversibility method leads to optimal controls which are both feasible and highly accurate. Shown in Figure~\ref{fig:optimal_pulses} are two sets of optimal controls for a 4-transmon system. These controls were initialized with a constant initial condition and over 1000 optimization iterations achieved infidelities below $10^{-4}$ for a $100$ ns control time. 
\par
While these fidelities neglect decoherence, they are much better than state-of-the-art two-transmon operations and are on a similar time scale of two-transmon operations in real devices \cite{krantz2019quantum}. Additionally, the optimized controls are extremely smooth and have well-defined amplitude both of which are within current experimental limitations \cite{krantz2019quantum}.
\par
As an additional validation, we visualize in Figure~\ref{fig:convergence} the convergence of infidelity with increasing optimization iterations. Similar to~\cite{PhysRevA.95.042318,10.1007/978-3-031-08760-8_11} we use the ADAM optimizer with a learning rate of $10^{-3}$. We find that there are only small differences between the convergence of the optimizer with the reversibility method compared to the naive JAX method. This is to be expected as numerical errors due to imperfect reversibility begin to propagate through the derivative calculation and will therefore drive subtle differences in convergence.

\begin{figure}[h]
\begin{center}
    \includegraphics[scale=0.7]{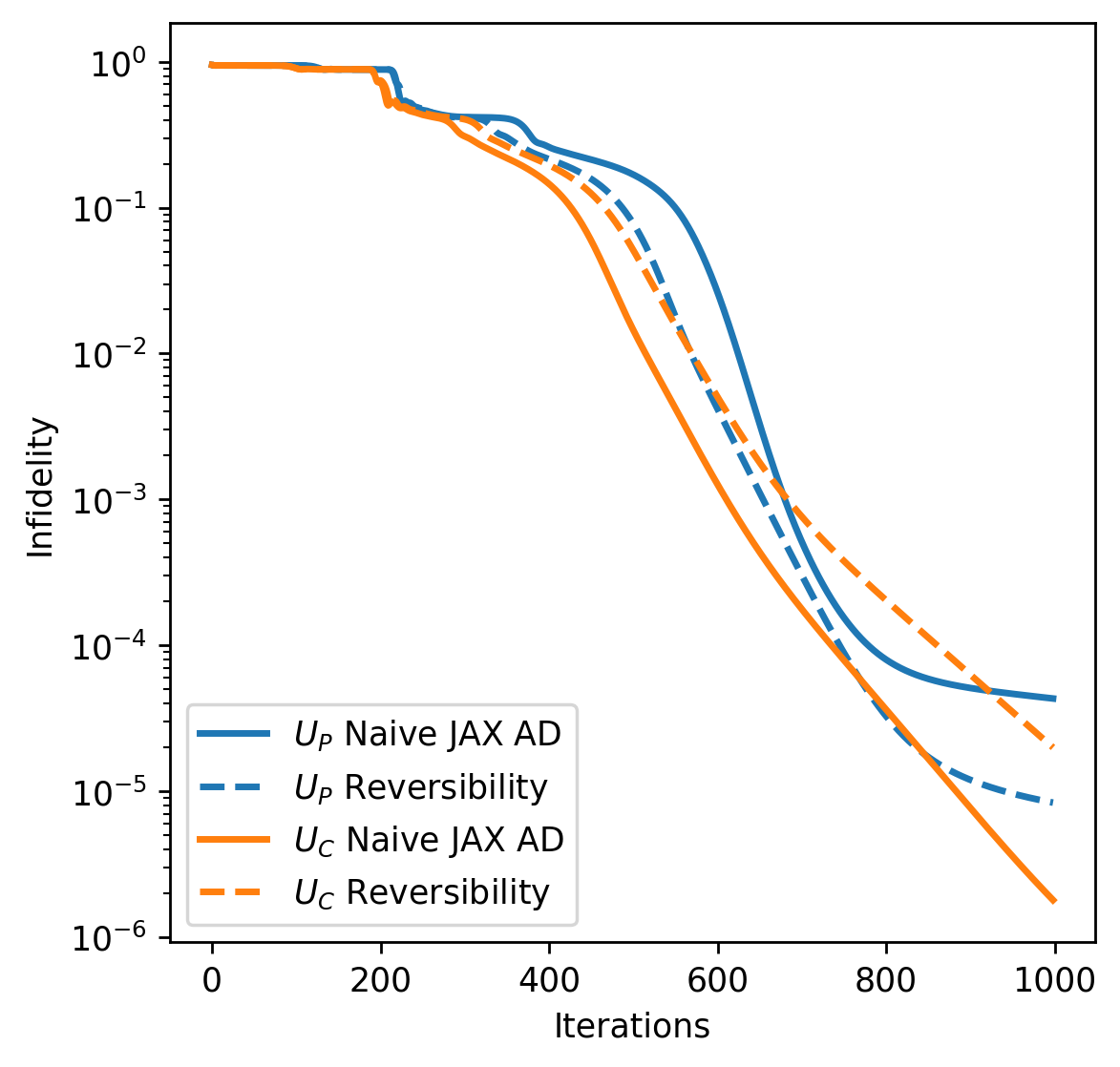}
    \end{center}
    \caption{The convergence to a set of optimal controls for two different target unitaries $U_P(t/q)$ and $U_C(t/q)$ and two different AD techniques. Each simulation uses the same device Hamiltonian and assumes $t/q = 0.01$ ns, $J=1$, $V=1$.}
    \label{fig:convergence}
\end{figure}

We explored the performance and memory requirements of the naive AD, reversibility, checkpointing, and checkpointing with reversibility approaches in three sets of experiments. In the first 
set, we vary the size of the lattice, thereby varying the number of qubits in the system. In the second set, we vary the number of timesteps 
in each iteration of the optimization process. Finally, we vary the interval between checkpoints for the checkpointing and the checkpointing 
with reversibility approach. Our experiments were conducted on a cluster where each compute node was connected to 8 NVIDIA A100 40GB GPUs. Each node contained
1TB DDR4 memory and 320GB GPU memory. We report the time taken to execute 20 iterations of the optimization procedure.
We used the JAX memory profiling capability in conjunction with {\tt GO pprof} to measure the memory needs for each case.

\noindent {\bf Vary Qubits} In these experiments, we fixed the width of the lattice to two and varied the length of the lattice. 
The results in Figure~\ref{fig:vary_qubits} (right) 
show that the device memory requirements for the standard approach are highest whereas the requirements for reversibility are lowest.  
We note that the standard approach can be executed at most for a $2\times3$ lattice made up of $7$ qubits and runs out of available device memory thereafter. The periodic checkpointing approach and reversibility approaches can be run for at most a $2\times4$ lattice made up of $10$ qubits and run out of available device memory thereafter. Figure~\ref{fig:vary_qubits} (left) also shows the execution time for the various approaches. 

\begin{figure} [h]
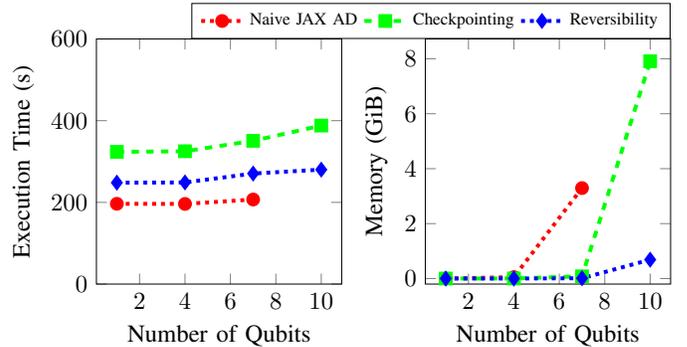

\begin{center}
    \includestandalone[width=1.0\linewidth]{figures_edge/vary_qubits_combined}
    \end{center} \vspace{-2em}
    \caption{Comparison of execution time and device memory requirements for standard AD, periodic checkpointing, and full reversibility with increasing number of qubits in the lattice. The QOC simulations consisted of $N=500$ time steps with a checkpoint period of $C=\lfloor \sqrt{N}\rfloor=22$. \label{fig:vary_qubits}}
\end{figure}

\noindent {\bf Vary Timesteps}
Next, we fixed the size of the lattice to $2\times3$ made up of $7$ qubits and varied the number of time steps, $N$. For periodic checkpointing, we used the optimal checkpoint period, $C=\lfloor\sqrt{N}\rfloor$. Our results are consistent with ~\cite{10.1007/978-3-031-08760-8_11}. The time is roughly linear in $N$ and independent of $C$. Periodic checkpointing and full reversibility are slower than naive AD. Full reversibility is somewhat faster than periodic checkpointing. The memory requirements of naive AD rise rapidly with more timesteps, while reversibility and checkpointing do not rise appreciably.

\begin{figure} [h]
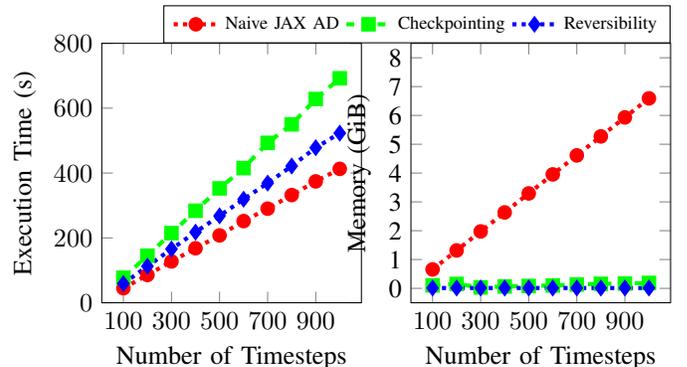

\begin{center}
    \includestandalone[width=1.0\linewidth]{figures_edge/vary_steps_combined}
    \end{center} \vspace{-2em}
    \caption{Comparison of the execution time and device memory requirements for standard AD, periodic checkpointing, and  full reversibility approaches with increasing number of time steps. The QOC simulation consisted of $7$ qubits. The checkpoint period was chosen to be the square root of the number of time steps. \label{fig:vary_steps_time}}
\end{figure}

\noindent {\bf Vary Checkpoints}
We examined the dependence of execution time and memory requirements on the checkpointing period, $C$, keeping the size of the lattice fixed at $2\times3$ made up of $7$ qubits and the number of time steps fixed at $N=500$. Again the results obtained in in Figure~\ref{fig:vary_checks} are consistent with ~\cite{10.1007/978-3-031-08760-8_11}. The time taken is roughly independent of $C$. Periodic checkpointing with reversibility is somewhat faster than periodic checkpointing alone. The memory requirements of periodic checkpointing with reversibility vary as a function of $\frac{N}{C}$. The memory requirements of periodic checkpointing alone vary as a function of $\frac{N}{C}+C$.

\begin{figure} [h]
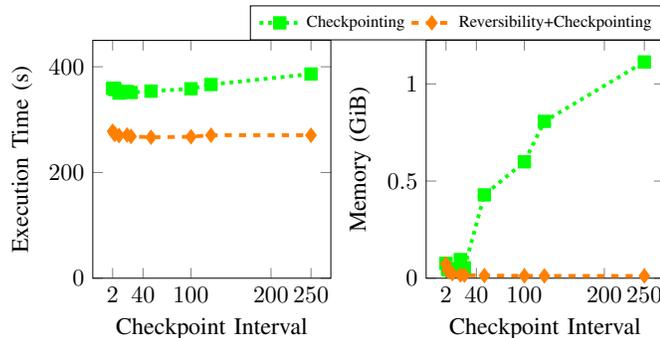

\begin{center}
    \includestandalone[width=1.0\linewidth]{figures_edge/vary_checks_combined}
\end{center} \vspace{-2em}
    \caption{Comparison of the execution time and device memory requirements for  periodic checkpointing and checkpointing plus reversibility approaches with increasing number of time steps. The QOC simulations consisted of $500$ time steps and $7$ qubits. 
    }
    \label{fig:vary_checks}
\end{figure}
\section{Conclusion}
\label{sec:conclusion}
We have demonstrated the application of three advanced AD approaches, implemented in the JAX differentiable programming framework, to lattice gauge theory. These approaches increase the number of qubits that can be simulated by reducing the memory requirements of automatic differentiation.

\bibliographystyle{splncs04}
\bibliography{qocjax1}


\end{document}

%% file: figures_edge/vary_qubits_combined.tex
\begin{tikzpicture}
   \begin{axis}[
                xlabel = {Number of Qubits},
                ylabel = { Execution Time (s)}, 
                 xtick={2,4,6,8,10},
                ymin=0,
                ymax=600,
                mark options={solid},
                width=10em,
                height=10em,
                scale only axis,
                    ]
      \addplot[ultra thick, red, dotted, mark=*] coordinates {
(	1	,	196.542	)
(	4	,	196.2094	)
(	7	,	206.9868	)
      };
      \addplot[ultra thick, green, dashed, mark=square*] coordinates {
(	1	,	323.6796	)
(	4	,	325.1791	)
(	7	,	350.6134	)
(	10	,	388.0041	)
      };
      \addplot[ultra thick, blue, dotted, mark=diamond*] coordinates {
(	1	,	248.2245	)
(	4	,	248.7652	)
(	7	,	270.5822	)
(	10	,	280.1378	)
      };
   \end{axis}
\end{tikzpicture}\hspace{-6.5em}
\begin{tikzpicture}
   \begin{axis}[
                xlabel = {Number of Qubits},
                ylabel = {Memory (GiB)}, 
                xtick={2,4,6,8,10,12},
                ymin=-0.2,
                mark options={solid},
                width=10em,
                height=10em,
                scale only axis,
                legend style={
                    at={(1.0,1.0)},
                    anchor=south east,
                    legend columns=3}
                    ]
      \addplot[ultra thick, red, dotted, mark=*] coordinates {
(	1	,	0.001570158	)
(	4	,	0.052492008	)
(	7	,	3.293515501	)
      };
      \addplot[ultra thick, green, dashed, mark=square*] coordinates {
(	1	,	0.001303129	)
(	4	,	0.002826099	)
(	7	,	0.081789589	)
(	10	,	7.910195789	)
      };
      \addplot[ultra thick, blue, dotted, mark=diamond*] coordinates {
(	1	,	0.00093154	)
(	4	,	0.001345272	)
(	7	,	0.011310196	)
(	10	,	0.687654142	)
      };
      \legend{\scriptsize{Naive JAX AD},
              \scriptsize{Checkpointing},
              \scriptsize{Reversibility},
        }
   \end{axis}
\end{tikzpicture}

%% file: figures_edge/vary_steps_combined.tex
\begin{tikzpicture}
   \begin{axis}[
                xlabel = {Number of Timesteps},
                ylabel = {Execution Time (s)}, 
                xtick={100,300,500,700,900},
                ymin=0,
                ymax=800,
                mark options={solid},
                width=10em,
                height=10em,
                scale only axis,
                legend style={
                    at={(0.25,1.0)},
                    anchor=north,
                    legend columns=1}
                    ]
      \addplot[ultra thick, red, dotted, mark=*] coordinates {
	(	100	,	44.8912	)
	(	200	,	85.6705	)
	(	300	,	127.4465	)
	(	400	,	167.9695	)
	(	500	,	207.6426	)
	(	600	,	251.8268	)
	(	700	,	290.1174	)
	(	800	,	332.1035	)
	(	900	,	374.2475	)
	(	1000	,	412.8788	)
      };
      \addplot[ultra thick, green, dashed, mark=square*] coordinates {
	(	100	,	77.3194	)
	(	200	,	145.1751	)
	(	300	,	214.6819	)
	(	400	,	283.7474	)
	(	500	,	352.6372	)
	(	600	,	415.2924	)
	(	700	,	493.0372	)
	(	800	,	550.09	)
	(	900	,	628.285	)
	(	1000	,	692.1324	)
      };
      \addplot[ultra thick, blue, dotted, mark=diamond*] coordinates {
	(	100	,	59.2331	)
	(	200	,	112.5251	)
	(	300	,	165.6161	)
	(	400	,	217.7419	)
	(	500	,	267.9974	)
	(	600	,	319.2797	)
	(	700	,	368.8924	)
	(	800	,	421.5968	)
	(	900	,	478.1153	)
	(	1000	,	523.0016	)
      };
   \end{axis}
\end{tikzpicture}\hspace{-8em}
\begin{tikzpicture}
   \begin{axis}[
                xlabel = {Number of Timesteps},
                ylabel = {Memory (GiB)}, 
                xtick={100,300,500,700,900},
                ymin=-0.5,
                mark options={solid},
                width=10em,
                height=10em,
                scale only axis,
                ymax=8.5,
                ytick={0,1,2,3,4,5,6,7,8},
                yticklabels={0,1,2,3,4,5,6,7,8},
                legend style={
                    at={(1.0,1.0)},
                    anchor=south east,
                    legend columns=3}
                    ]
      \addplot[ultra thick, red, dotted, mark=*] coordinates {
(	100	,	0.655773039	)
(	200	,	1.315208654	)
(	300	,	1.97464427	)
(	400	,	2.634079885	)
(	500	,	3.293515501	)
(	600	,	3.952951117	)
(	700	,	4.612386732	)
(	800	,	5.271822348	)
(	900	,	5.931257963	)
(	1000	,	6.590693579	)
      };
      \addplot[ultra thick, green, dashed, mark=square*] coordinates {
(	100	,	0.101839991	)
(	200	,	0.152166824	)
(	300	,	0.032385216	)
(	400	,	0.065300093	)
(	500	,	0.081789589	)
(	600	,	0.099995012	)
(	700	,	0.132907858	)
(	800	,	0.156020489	)
(	900	,	0.166138353	)
(	1000	,	0.18704689	)
      };
      \addplot[ultra thick, blue, dotted, mark=diamond*] coordinates {
(	100	,	0.011239986	)
(	200	,	0.011257839	)
(	300	,	0.011275444	)
(	400	,	0.01129282	)
(	500	,	0.011310196	)
(	600	,	0.011328526	)
(	700	,	0.011345663	)
(	800	,	0.011363277	)
(	900	,	0.011381245	)
(	1000	,	0.011398382	)
      };
      \legend{\scriptsize{Naive JAX AD},
              \scriptsize{Checkpointing},
              \scriptsize{Reversibility},
        }
   \end{axis}
\end{tikzpicture}

%% file: figures_edge/vary_checks_combined.tex
\begin{tikzpicture}
   \begin{axis}[
                xlabel = {Checkpoint Interval},
                ylabel = { Execution Time (s)},
                xtick={2,40, 100, 200, 250, 500},
                ymin=0,
                ymax=450,
                width=10em,
                height=10em,
                scale only axis,
                mark options={solid},
                legend style={
                    at={(0.25, 0.0)},
                    anchor=south,
                    legend columns=1}
                    ]
      \addplot[ultra thick, green, dotted, mark=square*] coordinates {
	(	2	,	359.4474	)
	(	4	,	359.5221	)
	(	5	,	356.8579	)
	(	10	,	350.2352	)
	(	20	,	353.276	)
	(	25	,	351.3451	)
	(	50	,	353.847	)
	(	100	,	358.3551	)
	(	125	,	366.3843	)
	(	250	,	386.3046	)
      };
      \addplot[ultra thick, orange, dashed, mark=diamond*] coordinates {
	(	2	,	278.0508	)
	(	4	,	272.7858	)
	(	5	,	272.6289	)
	(	10	,	269.88	)
	(	20	,	270.9288	)
	(	25	,	268.2368	)
	(	50	,	266.6507	)
	(	100	,	267.6116	)
	(	125	,	270.5553	)
	(	250	,	270.4935	) 
      };
   \end{axis}
\end{tikzpicture}\hspace{-4em}
\begin{tikzpicture}
   \begin{axis}[
                xlabel = {Checkpoint Interval},
                ylabel = { Memory (GiB)},
                xtick={2,40, 100, 200, 250, 500},
                ymin=0,
                mark options={solid},
                width=10em,
                height=10em,
                scale only axis,
                legend style={
                    at={(1.0,1.0)},
                    anchor=south east,
                    legend columns=2}
                    ]
      \addplot[ultra thick, green, dotted, mark=square*] coordinates {
(	2	,	0.07626174	)
(	4	,	0.048241463	)
(	5	,	0.043478642	)
(	10	,	0.038398848	)
(	20	,	0.095992289	)
(	25	,	0.053132782	)
(	50	,	0.429016914	)
(	100	,	0.599039993	)
(	125	,	0.807123537	)
(	250	,	1.112996511	)
      };
      \addplot[ultra thick, orange, dashed, mark=diamond*] coordinates {
(	2	,	0.073255816	)
(	4	,	0.04228301	)
(	5	,	0.036038485	)
(	10	,	0.023549452	)
(	20	,	0.017304935	)
(	25	,	0.016056032	)
(	50	,	0.013558226	)
(	100	,	0.012309322	)
(	125	,	0.012059536	)
(	250	,	0.011559982	)
      };
      \legend{\scriptsize{Checkpointing},
              \scriptsize{Reversibility+Checkpointing},
        }
   \end{axis}
\end{tikzpicture}